# Operationalizing the AAPT Learning Goals for the Lab

N.G. Holmes and Emily M. Smith
Laboratory of Atomic and Solid State Physics, Cornell University

Calls for reform to instructional labs means many instructors and departments are facing the daunting task of identifying goals for their introductory lab courses. Fortunately, the American Association of Physics Teachers (AAPT) released a set of recommendations for learning goals for the lab to support lab redevelopment [1]. Here we outline the process we have undergone to identify a set of learning goals for the labs that operationalize those provided by the AAPT. We also provide two examples of newly developed lab activities that aim to meet those goals to demonstrate this operationalization. We aim to provide departments and instructors with a few ideas of a procedure that they can follow or a set of goals that they can use to align lab instruction with the AAPT learning goals.

## 1. WHY START WITH LEARNING GOALS?

Developing learning goals is a critical part of any course transformation. Many instructional transformations use a backwards course design, where design starts at the end: what do I want my students to get out of this course? Strategies such as the Course Transformation Model used by the Science Education Initiatives at the University of Colorado-Boulder and the University of British Columbia have been found to result in improved student learning outcomes [2]. The process starts with identifying what you want students to be able to do by the end of the course (goals), follows with determining how to measure whether students have achieved those goals (assessment), and culminated in identifying ways to support students in achieving those goals (instructional strategies).

Effective learning goals are described as being SMART: **S**pecific, **M**easurable, **A**ttainable, **R**elevant, and **T**ime-limited. SMART learning goals help make a course transparent. They help instructors make deliberate choices when designing classroom activities and help students know what they need to know [3].

## 2. AAPT GOALS FOR THE LAB

In 2014, the AAPT released a set of recommendations for learning goals for the full undergraduate physics lab curriculum [1]. The recommendations span a broad array of goals that fall into six themes (Figure 1):

**Modeling** includes defining models as abstract representations of physical and measurement systems with limitations and approximations, and developing, evaluating, or testing such models.

**Designing experiments** includes developing, evaluating, and troubleshooting experiments to test models.

**Developing technical and practical skills** includes developing a range of skills related to experimentation such as working with specific equipment (such as oscilloscopes).

**Analyzing and visualizing data** includes understanding and implementing a range of statistical and graphical methods to evaluate and interpret data and their uncertainties.

**Constructing knowledge** includes the process of using data to generate ideas and conclusions about the physical world.

**Communicating physics** includes argumentation from evidence and synthesizing experimental methods and outcomes for broad consumption.

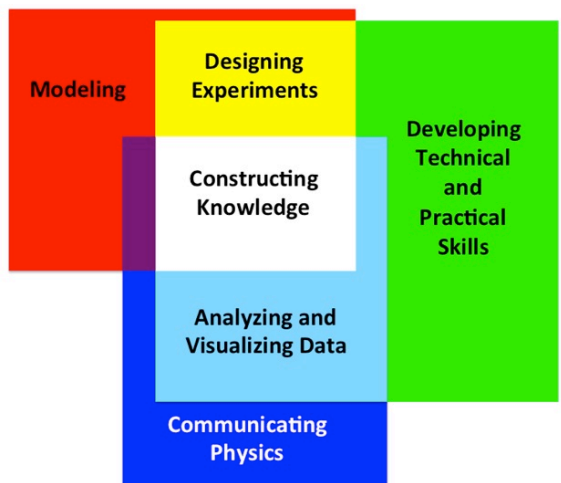

Figure 1. The six themes of learning goals recommended by the American Association of Physics Teachers from [1]

These goals are thorough descriptions of what physics curricula could aim to achieve through their lab instruction, spanning introductory and advanced lab courses. Turning these recommendations into SMART learning goals for a particular course can be tricky. Where do you start? How do you narrow down all the possibilities into an attainable set of specific goals? What's measurable? Here we describe our process as one example of how to make use of this great resource.

### 3. OPERATIONALIZING FOR THE INTRO LAB

We have been using the course transformation model and the AAPT recommendations to redevelop the labs for our calculus-based introductory physics sequences. In all courses, the labs are a required part of the lecture course (not stand-alone lab courses). The transformation began with the physics majors sequence, followed by the larger sequence intended for engineering majors.

In such a transformation, it is important to solicit input from departmental faculty and stakeholders to identify what's important and *relevant* [2,4]. We first met with faculty in the department through one-on-one meetings and focus groups. The discussions were relatively unstructured, but always included one key question: What do you wish students entering the advanced lab or your research group were already able to do? Faculty and instructors had long lists of very specific, foundational activities they wished students could do. An unstructured discussion was used deliberately to capture a diverse set of possible goals not influenced by the interviewer. Subsequent questions probed the interviewees' responses to this key question, which were necessary to understand and clarify the interviewees' priorities.

We compiled an extensive list of possible goals from the contents of these discussions. We categorized the goals into themes and began to prioritize the list according to what came up the most frequently. The categorized and ordered list was discussed with a committee of faculty and instructors, some of whom had been interviewed. Following the discussion, we prioritized what needed to be covered in the first year (*attainable*) and removed items that we decided could be saved until the sophomore or junior labs (*time-limited*) as well as ones that were only mentioned by one interviewee (*relevance*).

> By the end of this activity, you should be able to:
> 1. Collect data and revise an experimental procedure iteratively and reflectively,[M, D, T]
> 2. Evaluate the process and outcomes of an experiment quantitatively and qualitatively,[M, A]
> 3. Extend the scope of an investigation whether or not results come out as expected,[M, D, K]
> 4. Communicate the process and outcomes of an experiment,[C] and
> 5. Conduct an experiment collaboratively and ethically.[K, C]

Next, we created a concept map of the goals to identify how they connect: What concepts build on other concepts? Which ones need to be introduced first (have multiple dependencies)? From this, we developed a rough time sequence for instruction and made sure there were no isolated goals. Language was carefully crafted to ensure goals were also *specific* and *measurable*. The final set of learning goals, which fell under five headings (Fig. 2), were further vetted by the faculty committee. Under each main learning goal, there are a series of more specific learning objectives (see [5] for the full list).

*Figure 2 Learning goals for the introductory physics lab sequence at Cornell University. Alignment with the AAPT goal headings are indicated with superscripts. M=Modeling, D=Designing experiments, T=Technical skills, A=Analyzing data, K=constructing Knowledge, C=Communicating physics.*

The list clearly targets goals for physics majors and our key interview question deliberately focused on skills related to physics research. Our course transformation, however, will include both majors and non-majors courses. We have discussed the learning goals with curriculum leaders in our engineering departments and they are fully supportive of this list of goals, many of which reflect goals of those departments as well.

We next provide two examples of activities to meet some of the specific sub-goals associated with Fig. 2. Both examples come from our introductory, calculus-based physics courses (first-semester mechanics and second-semester electricity and magnetism) and both build from activities that were already being carried out in our traditional labs (to save costs). We provide these as a demonstration of the operationalization of the goals, rather than a coherent description of the course transformation process.

4. **EXAMPLE ACTIVITY: MECHANICS**

Throughout the mechanics course, we prioritized the learning objectives in the first three categories in Fig. 2. The example activity took place during the fourth and fifth session of the course. The full set of sub-goals for the lab are in Fig. 3.

By this point in the course, we wanted to reinforce goals introduced in the first few weeks, particularly related to revising experimental procedures iteratively and reflectively (Goal #1 in Fig. 2). New goals focused on using fitting procedures to evaluate data quantitatively (Goal #2 in Fig. 2).

### A. Choosing an activity

To achieve these goals, we needed: an activity that involved two continuous and linearly-related variables; two reasonable models that could be compared; relatively simple data collection so that students could focus on the analysis and iterate; and relatively messy data that would warrant iteration and reflection.

We chose a lab where students drop stacks of coffee filters, use a position sensor to identify regions in which the stacks travel at terminal velocity, and then evaluate whether the terminal velocity is linear or quadratic with the mass of the stack.

> By the end of this activity, you should be able to:
> - Identify, minimize and/or quantify sources of statistical uncertainty, systematics, or mistakes, [D, A]
> - Decide what and how much data are to be gathered to produce reliable measurements, [D]
> - Describe how weighted least-squares provides a measure of the best fit, [A]
> - Plot a residual graph and calculate the chi-squared statistic to compare data to a model, [A]
> - Design, carry out, and improve an experiment to test competing physical models. [M, D, K]

*Figure 3 Sub-goals for a mid-semester mechanics lab activity (first semester). Alignment with the AAPT goal headings are indicated with superscripts.*

### B. How does the lab achieve each sub-goal?

Throughout the activity, students are given significant control (agency) over the experimentation process and make many of the related decisions.

The first two sub-goals about uncertainty are achieved through the many decisions involved with measuring terminal velocity of falling coffee filters. Students must decide what range of position versus time data is sufficiently linear to be reasonably confident that the filters reached terminal velocity and how to quantify their uncertainty. They must also decide what 'sufficient' or 'reasonable' mean in these contexts, with feedback from the instructor. Questions include: How many trials are sufficient given the time constraints? From what height should the filters be dropped?

For teaching students about fitting (the third and fourth goals), we use an invention task [6,7]. An invention task provides students with a problem to solve prior to receiving instruction on the canonical solution. They do not need to invent the canonical solution, as they receive the necessary information in a follow-up class discussion. The activity that we used can be found at [7]. After the activity, students use the new tools in the analysis of their terminal velocity data to compare the two possible models.

For the final goal, we once again rely on the features of the terminal velocity data. The two possible models look very similar if a) the uncertainty is too large or b) the range of masses is too small. This means students must improve their experiment to distinguish the two models, and it provides some fairly straightforward ways to do so (more trials, different drop heights, more careful collection or selection of data, more masses). This also helps reinforce the big ideas about fitting: quantitatively, one model will have a smaller chi-squared statistic than the other, but the plots (raw data and residuals) raise questions as to the practical significance of that result.

## 5. EXAMPLE ACTIVITY: E&M

In our electricity and magnetism course, we continue to build on the learning objectives that fall under the first three categories in Fig. 2 and prioritize the development of collaboration skills and ethical experimentation practices, the fifth category. At the beginning of the semester, we wanted to review the learning goals from the previous semester, while also introducing an explicit focus on collaboration. The full set of sub-goals for the lab are in Fig. 4.

---

By the end of this activity, you should be able to:
- Generate a model to test by consulting previous data and results and/or theory and using that model to make predictions about expected measurements, data, and results, [M, K]
- When data and results do not come out as expected:
  – Test whether the results are reproducible under the same conditions or with improved precision, [M, D, K] and/or
  – Design new experiments/tests to explore other explanations for the disagreement, [M, D, K]
- When data and results do come out as expected:
  – Test whether the results hold with higher levels of accuracy and precision, and/or [M, D, K]
  – Extend the scope of the experiment to check if there is "new" physics at these levels, [M, D, K]
- Brainstorm with group members to construct a diverse set of ideas when making decisions, [C] and
- Share experimentation responsibilities with group members. [C]

---

Figure 4 Sub-goals for the first E&M lab activity (second semester). Alignment with the AAPT goal headings are indicated with superscripts.

### A. Choosing an activity

To achieve these goals, we needed a lab activity with relatively simple analyses that would incorporate opportunities to design and extend experiments to test models. It also needed to provide sufficient ambiguity that would necessitate within-group and across-group collaboration. We used an activity where students generate models of electrostatics from a series of demonstrations (e.g. electroscopes, attraction of packing peanuts, simple Faraday cages), and then design experiments to test those models.

### B. How does the lab achieve each sub-goal?

In this lab, the instructor shows each group a different electrostatics demonstration without explanation of any aspect of the set-up. To practice the first three goals in Fig. 4, students work in their groups to brainstorm models that might explain the demonstration and design ways to test those models. They then extend their investigation by using their model to make predictions about the outcomes of other groups' demonstrations, testing those predictions, and refining their model. This sequence of events is inspired by the framework from the Investigative Science Learning Environments [8]. In each case, the demonstrations and set up are provided, but both can be fully manipulated, and the experimental design goals focus on model testing through controlling variables (different from the data acquisition goals in the mechanics example).

To practice the collaboration goals, students rotate though group roles similar to those of cooperative groups [9] but that emphasize roles in experimental physics (a principal investigator, a science communicator, and Reviewer #2). By formally rotating students through roles within the lab, they take on different perspectives and responsibilities in the experimentation process. By working together to develop and test models, groups must construct several testable ideas. As they extend to new demonstrations, they are faced with several new situations where they must brainstorm new testable ideas. Partway through the session, the instructor pauses the class for a "mini-conference" where groups share their results and models. Through this discussion, students hear diverse ideas from classmates that they incorporate moving forward. It also leads to additional collaboration between groups for the second half of the lab.

6. SUMMARY

Backwards-course design and the Course Transformation Model begin with defining SMART learning goals. The recommendations from the AAPT have provided a framework for departments to generate their own SMART learning goals for their laboratory curriculum. This article presents an example of such a process, as well as how to align existing lab activities with those learning goals. For more examples of lab activities tied to these goals, please visit cperl.lassp.cornell.edu/.

ACKNOWLEDGEMENTS

We would like to thank Carl Wieman and Peter Lepage for feedback on the manuscript and all the faculty in the physics department at Cornell that participated in the interviews and focus groups. This work is partially supported by the Cornell University College of Arts and Sciences Active Learning Initiative.